# Integrated Routing Protocol for Opportunistic Networks

Anshul Verma
Computer Science and Engineering Dept.
ABV-Indian Institute of Information Technology and Management, Gwalior, India
E-mail: anshulverma87@gmail.com

Dr. Anurag Srivastava
Applied Science Dept.
ABV-Indian Institute of Information Technology and Management, Gwalior, India
E-mail: anurags@iiitm.ac.in

*Abstract*—In opportunistic networks the existence of a simultaneous path is not assumed to transmit a message between a sender and a receiver. Information about the context in which the users communicate is a key piece of knowledge to design efficient routing protocols in opportunistic networks. But this kind of information is not always available. When users are very isolated, context information cannot be distributed, and cannot be used for taking efficient routing decisions. In such cases, context oblivious based schemes are only way to enable communication between users. As soon as users become more social, context data spreads in the network, and context based routing becomes an efficient solution. In this paper we design an integrated routing protocol that is able to use context data as soon as it becomes available and falls back to dissemination-based routing when context information is not available. Then, we provide a comparison between Epidemic and PROPHET, these are representative of context oblivious and context aware routing protocols. Our results show that integrated routing protocol is able to provide better result in term of message delivery probability and message delay in both cases when context information about users is available or not.

## I. INTRODUCTION

The opportunistic network is an extension of Mobile Ad hoc Network (MANET). Wireless networks' properties, such as disconnection of nodes, network partitions, mobility of users and links' instability, are seen as exceptions in traditional network. This makes the design of MANET significantly more difficult [1].

Opportunistic networks [2] are created out of mobile devices carried by people, without relying on any preexisting network topology. Opportunistic networks consider disconnections, mobility, partitions, etc. as norms instead of the exceptions. In opportunistic network mobility is used as a technique to provide communication between disconnected 'groups' of nodes, rather than a drawback to be solved.

In opportunistic networking a complete path between two nodes wishing to communicate is unavailable [3]. Opportunistic networking tries to solve this problem by removing the assumption of physical end-to-end connectivity and allows such nodes to exchange messages. By using the store-carry-and-forward paradigm [4] intermediate nodes store messages when there is no forwarding opportunity towards the destination, and exploit any future contact opportunity with other mobile devices to bring the messages closer and closer to the destination.

Therefore routing is one of the most compelling challenges. The design of efficient routing protocols for opportunistic networks is generally a difficult task due to the absence of knowledge about the network topology. Routing performance depends on knowledge about the expected topology of the network [5]. Unfortunately, this kind of information is not always available. Context information is a key piece of knowledge to design efficient routing protocols. Context information represents users' working address and institution, the probability of meeting with other users or visiting particular places. It represents the current working environment and behavior of users. It is very help full to identify suitable forwarders based on context information about the destination. We can classify the main routing approaches proposed in the literature based on the amount of context information of users they exploit. Specifically, we identify two classes, corresponding to context-oblivious and context-aware protocols.

Protocols in Context-oblivious routing class as Epidemic Routing Protocol [6] are only solution when context information about users is not available. But they generate high overhead, network congestion and may suffer high contention. Context-based routing provides an effective congestion control mechanism and with respect to context-oblivious routing, provides acceptable QoS with lower overhead. Indeed, PRoPHET [7] is able to automatically learn the past communication opportunities determined by user's movement patterns and exploit them efficiently in future. This autonomic, self-learning feature is completely absent in Context-oblivious routing schemes. But context based routing protocols provide high overhead, message delay and less success full message in absence of context information about users. We have proved this by implementing epidemic and PROPHET routing protocols in presence and absence of context information. We found epidemic is better in absence of context information while PROPHET gives better result in presence of context information. Therefore I decided to combine feature of these both protocols into a single integrated routing protocol, which will perform better in both cases when context information about user is available or not.

This paper represents our integrated routing protocol, and evaluates it through simulations. The rest of the paper is organized as follows. Section 2 describes main routing protocols of context oblivious and context based routing class which are epidemic and PROPHET, and describe some related work. In section 3 our proposed scheme is presented. In section 4 the simulation setup is given for routing protocols. Comparison and Result of epidemic, PROPHET and integrated routing protocols can be found in section 5. Finally section 6 discusses conclusion and looks into future work.

## II. RELATED WORK

### A. Epidemic Routing

Epidemic Routing provides the final delivery of messages to random destinations with minimal assumptions of topology and connectivity of the network. Final message delivery depends on only periodic pair-wise connectivity between mobile devices. The Epidemic Routing protocol works on the theory of epidemic algorithms [6]. Each host maintains two buffers, one for storing messages that it has originated and second for messages that it is buffering on behalf of other hosts. Each mobile device stores a summary vector that contains a compact representation of messages currently stored in buffer.

When two hosts come into communication range of one another, they exchange their summary vectors. Each host also maintains a buffer to keep list of recently seen hosts to avoid redundant connections. Summary vector is not exchanged with mobile devices that have been seen within a predefine time period. After exchanging their summary vectors mobile devices compare summary vectors to determine which messages is missing. Then each mobile device requests copies of messages that it does not contain.

Each message has a unique identification number and a hop count. The message identifier is a unique 32-bit number. The hop count field determines the maximum number of intermediate nodes that a particular message can travel. When hop count is one, messages can only be delivered to their final destinations. Larger values for hop count will increase message delivery probability and reduce average delivery time, but will also increase total resource consumption in message delivery [6].

### B. PROPHET routing

PROPHET [7], a Probabilistic ROuting Protocol using History of Encounters and Transitivity makes use of observations that real users mostly move in a predictable fashion. If a user has visited a location several times before, there is more probability to visit that location again. PROPHET uses this information to improve routing performance.

To accomplish this, PROPHET maintains delivery predictability metric at every node. This metric represents message delivery probability of a host to a destination. PROPHET is similar to Epidemic Routing but it introduces a new concept of delivery predictability. Delivery predictably is the probability for a node to encounter a certain destination. When two nodes meet, they also exchange delivery predictability information with summary vectors. This information is used to update the delivery predictability information of metric.

When a message comes at a node, node checks that destination is available or not. If destination is not available, node stores the message and upon each encounters with another device, it takes decision whether or not to transfer a message. Message is transferred to the other node if the other node has higher message delivery probability to the destination [7].

### C. Other work

Since routing is one of the most challenging issues in opportunistic networks, many researchers are working in this area. In this Section we are only mentioning some specific routings, which are representative of both context oblivious and context based routing protocols in opportunistic networks. The reader can also find a brief discussion on routing protocols for opportunistic networks in [8].

Context oblivious based algorithms also include network-coding-based routing [9]. In general, network coding-based routing reduces flooding, as it is able to deliver the same amount of information with fewer messages injected into the network [10].

Spray and Wait [11] routing provides a drastic way to reducing the overhead of Epidemic. Message is delivered in two steps: the spray phase and the wait phase. During the spray phase, source node and first receivers of the message spread multiple copies of the same message over the network. Then, in the wait phase each relay node stores its copy and eventually delivers it to the destination when it comes within reach.

Frequency of meetings between nodes and frequency of visits to specific physical places is used by MV [12] and MaxProp [13] as context information.

MobySpace routing [14] uses the mobility pattern of nodes as context information. The protocol uses a multi dimensional Euclidean space, named MobySpace, where possible contact between couples of nodes are represented by each axis and the probability of that contacts to occur are measured by the distance along axis. Two nodes that are close in the MobySpace, have similar sets of contacts. The best forwarding node for a message is the node that is as close as possible to the destination node in this space.

In Bubble Rap [15], social community users belong to is used as context information. Basically, Bubble Rap prefers nodes belonging to the same community of the destination as a good forwarder to this destination. If such nodes are not found, it forwards the message to the nodes, which have more chances of contact with the community of the destination. In Bubble Rap, communities are automatically detected via the patterns of contacts between nodes.

Other opportunistic routing protocols use the time lag from the last meeting with a destination as context information. Last Encounter routing [16] and Spray and Focus [17] routings are example of protocols exploiting such type of information.

Context-aware routing [18] uses an existing MANET routing protocol to connect nodes of the same MANET cloud. To transmit messages outside the cloud, a sender gives message to the node in its current cloud that has highest message delivery probability to the destination. This node waits to get in touch with destination or enters in destination's cloud with other nodes that has higher probability of meeting the destination. In context-aware routing context information is used to calculate probabilities only for those destinations each node is aware of.

With respect to context-aware routing, HiBOp is more general, HiBOp is a fully context-aware routing protocol completely described in [19]. HiBOp exploits every type of context information for taking routing decisions and also describes mechanism to handle this information. In HiBOp, devices share their own data when they come into contact with other devices, and thus learn the context they are immersed in. Nodes seem as good forwarders, which share more and more context data with the message destination.

## III. INTEGRATED ROUTING

Real users are likely to move around randomly or in predictable fashion, such that if a node has visited a location several times before, it is likely that it can visit that location again or can choose a new location that has never visited before. In this way users' movement can be predictable or unpredictable. We would like to make use of these observations to improve routing performance by combining probabilistic routing with flooding based routing and thus, we propose integrated routing protocol for opportunistic network.

To accomplish this, each node needs to know the contact probabilities to all other nodes currently available in the network. Every node maintains a probability matrix same as described in [7]. Each cell represents contact probability between to nodes x and y. Each node computes its contact probabilities with other nodes whenever the node comes in to contact with other nodes. Each node maintains a time attribute to other available nodes, the time attribute of a node is only updated when it meets with other nodes.

Two nodes exchange their contact probability matrices, when they meet. Nodes compare their own contact matrixes with other nodes. A node updates its matrix with another nodes' matrix if another node has more recent updated time attribute. In this way, two nodes will have identical contact probability matrices after communication.

### A. Probability calculation

The calculations of the delivery predictabilities have described in [7]. The first thing to do is to update the metric whenever a node meets with other nodes, so that nodes that are often met have a high message delivery probability. When node x meets node y, the delivery probability of node x for y is updated by (1).

$$P'_{xy} = P_{xy} + (1 - P_{xy}) P_0 \qquad (1)$$

Where $P_0$ is an initial probability, we used $P_0 = 0.75$. When node x does not meet with node y for some predefine time, the delivery probability decreases by (2).

$$P'_{xy} = \alpha^k P_{xy} \qquad (2)$$

Where α is the aging factor (a < 1), and k is the number of time units since the last update. When node x receives node y's delivery probabilities, node x may compute the transitive delivery probability through y to z by (3).

$$P'_{xz} = P_{xz} + (1 - P_{xz}) P_{xy} P_{yz} \beta \qquad (3)$$

Where β is a design parameter for the impact of transitivity, we used β = 0.25.

### B. Routing strategies

when a message arrives at a node, there might not be a path to the destination available so the node have to buffer the message and upon each encounters with another node, the decision must be made on whether or not to transfer a particular message. Furthermore, it may also be sensible to forward a message to multiple nodes to increase the probability that a message is really delivered to its destination.

Whenever a node meets with other nodes, they all exchange their messages (or as above, probability matrix). If the destination of a message is the receiver itself, the message is delivered. Otherwise, if the probability of delivering the message to its destination through this receiver node is greater than or equal to a certain threshold, the message is stored in the receiver's storage to forward to the destination. If the probability is less than the threshold, the receiver discards the message. If all neighbors of sender node have no knowledge about destination of message and sender has waited more than a configured time, sender will broadcast it to all its current neighbors. This process will be repeated at each node until it reaches to destination.

In this paper, we have developed a simple routing protocol –a message is transferred to the other node when two nodes meet, if the delivery probability to the destination of the message is higher than other node. But, taking these decisions is not an easy task. In some cases it might be sensible to select a fixed value and only give a message to nodes that have delivery probability greater than that fixed value for the destination of the message. On the other hand, when encountering a node with low delivery predictability, it is not certain that a node with a higher metric will be encountered within reasonable time. It may be possible destination is new and context information about destination is not spread in network. To solve these problems we introduce a new concept, our integrated routing distributes copies of message to all its neighbors same as flooding based techniques, after a configurable time, when node has not have any context information about destination of message.

Furthermore, we can also set the maximum number of copies of a message; a node can spread, to solve the problem of deciding how many nodes to give a certain message to. Distributing a message to a large number of nodes increases message delivery probability and decreases message delay, on the other hand, also increases resource consumption.

## IV. SIMULATION SETUP

We have currently implemented four different routing protocols epidemic, PROPHET, PROPHET (with no POIs) and integrated routing protocols in ONE (Opportunistic Network

Environment ) Simulator, all of which we consider in our evaluation. We are taking simulation scenario from [20], therefore we are not describing all things here. We are just showing here some important and new parameters.

We have used part of the Helsinki downtown area (4500×3400 m) as depicted in [20].For our simulations, we assume communication between modern mobile phones or similar devices. Devices has up to 20 MB of free RAM for buffering messages. Users travel on foot, in cars or trams. In addition, we have added to the map data some paths to parks, shopping malls and tram routes. We run our simulations with 100 nodes. Mobile nodes have different speed and pause time. Pedestrians move at random speeds of 0.5–1.5 m/s with pause times of 0–120 s. Cars are optional and, if present, make up 20% of the node count; they move at speeds of 10–50 km/h, pausing for 0–120 s. 0, 2, 4, or 6 trams run as speeds of 7–10 m/s and pause at each configured stop for 10–30 s. We assume Bluetooth (10 m range, 2 Mbit/s) and a low power use of 802.11b WLAN (30 m range, 4.5 Mbit/s). Mobile users (not the trams or throw-boxes) generate messages on average once per hour per node. We use message lifetimes of 3, 6, and 12 hours. We use message sizes uniformly distributed between 100 KB (text message) and 2 MB (digital photo).

Additionally, we define two scenarios POIs1 and POIs2 using different POIs each contains five groups and creates four POI groups (west containing 3, central 4, shops 22, and parks 11 POIs) [20]:

- **POIs1:** One node group runs MBM (map-based model), three choose their next destination with a probability p = 0.1 for each of the four POI groups, the last remaining one only chooses from the POI groups that are accessible by car otherwise a random target is selected.

- **POIs2:** We consider a preferred POI group for four of the node groups. A node chooses a POI with p = 0.4 from its preferred POI group, with p = 0.1 from each other POI group, and otherwise a random target.

## V. SIMULATION RESULTS

Now we compare the performance of epidemic, PROPHET and integrated routing protocols in both scenarios when context information is present or not. Here PROPHET (no POIs) stands for PROPHET routing protocol without context information about users, same meaning is here of integrated (no POIs) routing protocol. No POIs means, nodes have no information about destinations behavior and moving pattern, we do this by assigning 0.0 probabilities to each POI (point of interest). While, PROPHET stands for standard probabilistic routing protocol and integrated stands for our new routing protocol with context information about users.

Here figure 1 and 2 show message delay and message delivery probability of epidemic routing. Figure 3, 4, 5 and 6 represent message delay and message delivery probabilities of PROPHET and PROPHET with no POIs routing. Message delay and message delivery probabilities of integrated and integrated with no POIs are represented in figure 7, 8, 9 and 10.

Figure 11 show comparison of message delay between all routing protocols. It is cleared by this figure, when context information is available about users PROPHET gives minimum message delay probability 0.2370. But in absence of context information it gives maximum message delay probability 0.2824, that we represent by PROPHET (no POIs). Epidemic is totally flooding based routing protocol and does not require context information for message forwarding therefore it is not affected by unavailability of context information, and gives same message delay probability 0.2738 in both cases. Our own integrated routing gives 0.2480 and 0.2603 message delay probability in presence and absence of context information.

Comparison of message delivery probability between all routing protocols is shown in figure 12. Same as in case of message delay, PROPHET gives better message delivery probability 0.2981, but on unavailability of context information it gives worst message delivery probability 0.1978. Epidemic does not use context information, therefore gives same delivery probability 0.2334 in both cases. Our integrated routing gives 0.2822 delivery probability and 0.2506 delivery probability is given by integrated (no POIs) routing.

Table 1 shows a summary of message stats report of five routing protocols, which we have implemented. Here variable "sim_time" stands for total simulation time, "created, started, relayed, aborted and dropped" represent number of messages created by simulator, started for transmission, relayed by nodes and dropped by network. Whereas "delivery_prob" stands for total probability of messages delivery, "delay_prob" stands for total probability of messages delay, "hopcount_avg" represents average of intermediate nodes travelled by messages and "buffertime_avg" stands for Average of time Messages were buffered at nodes.

Our simulation results show that PROPHET gives batter result in presence of context information. When users are very isolated, context information cannot be distributed, and cannot be used for taking effective routing decisions. In this case PHROPHET gives worst result. Epidemic gives common result in both cases we have described above. And our integrated routing gives better result in both scenarios context information is available on not. Therefore integrated routing protocol is better when users are social or isolated.

## VI. CONCLUSIONS

In this work, we have proposed an integrated routing for opportunistic networks and evaluated its performance across a range of parameters' values, in comparison with Epidemic, PROPHET and PROPHET (with no POIs) routings. We have observed that our proposed integrated routing is able to meet out the challenges of other routing schemes for the

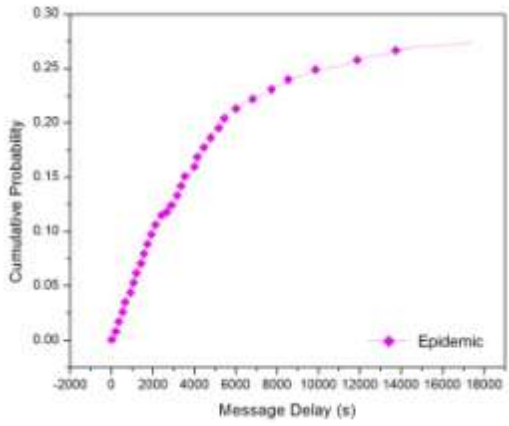

Figure 1. Message delay of Epidemic routing.

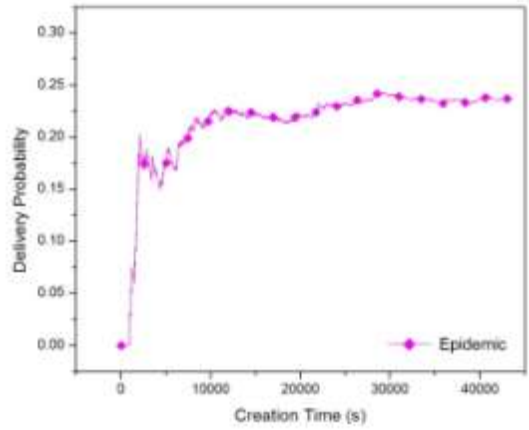

Figure 2. Delivery probability of Epidemic routing.

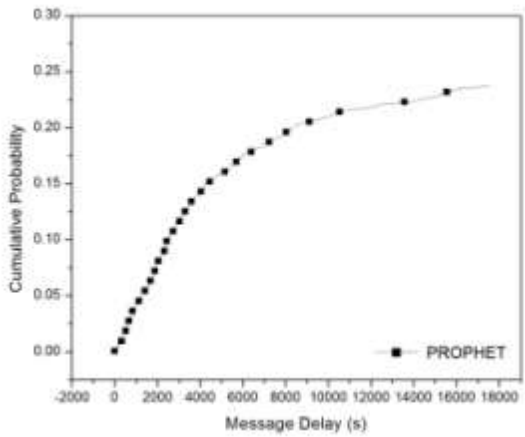

Figure 3. Message delay of PROPHET routing.

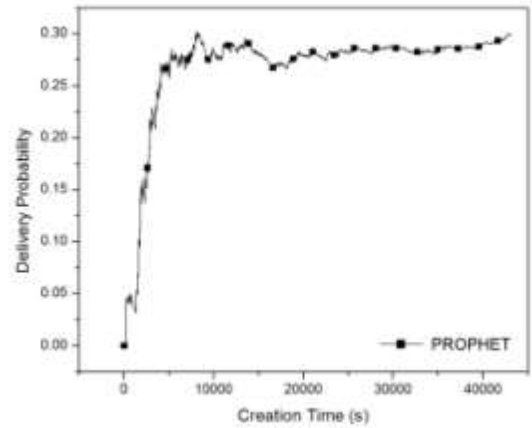

Figure 4. Delivery probability of PROPHET routing.

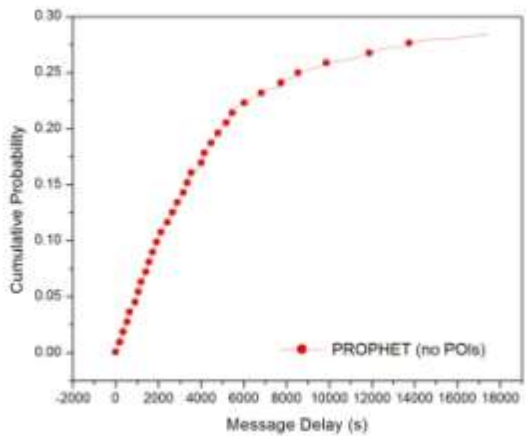

Figure 5. Message delay of PROPHET (no POIs) routing.

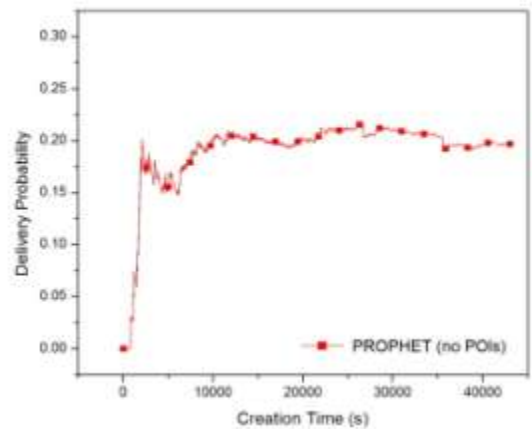

Figure 6. Delivery probability of PROPHET (no POIs) routing.

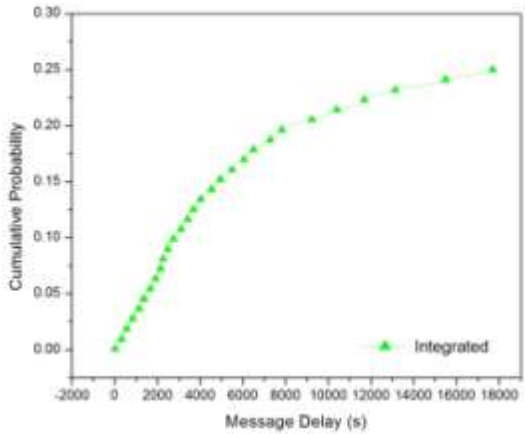

Figure 7. Message delay of Integrated routing.

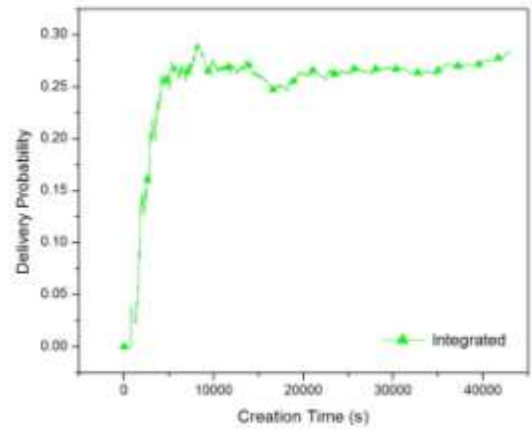

Figure 8. Delivery probability of Integrated routing.

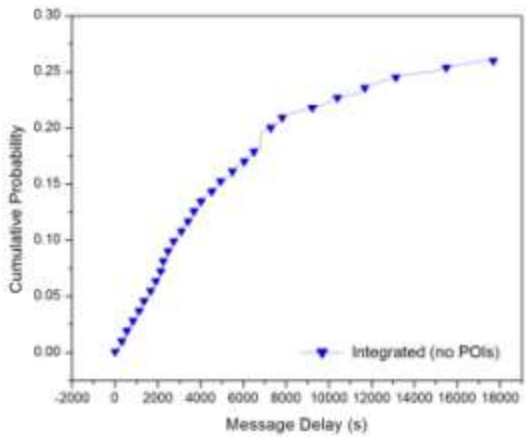

Figure 9. Message delay of Integrated (no POIs) routing.

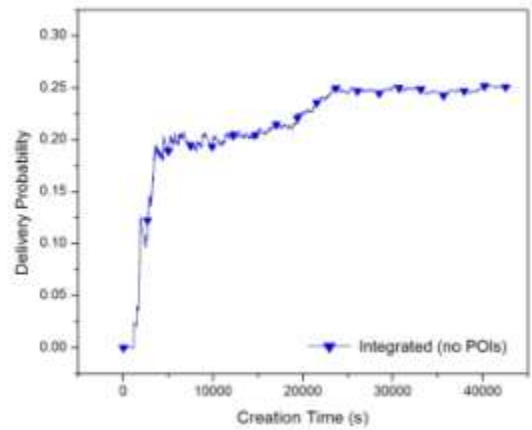

Figure 10. Delivery probability of Integrated (no POIs) routing.

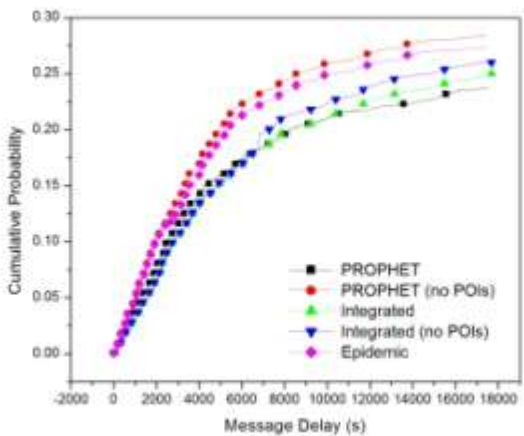

Figure 11. Comparison of Message delay of routing protocols.

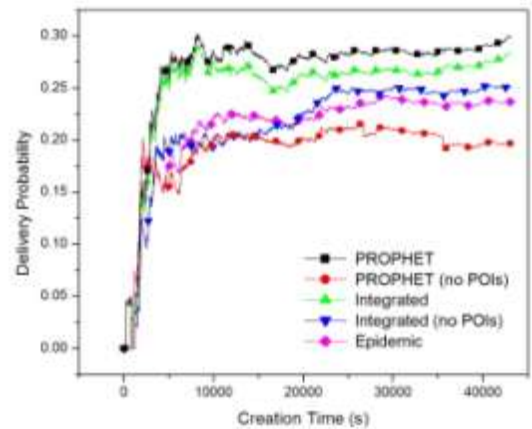

Figure 12. Comparison of Delivery probability of routing protocols.

TABLE I. MESSAGE STATS REPORT

| | Epidemic | PROPHET | PROPHET (no POI) | Integrated | Integrated (no POIs) |
|---|---|---|---|---|---|
| sim_time: | 43200.1000 | 43200.1000 | 43200.1000 | 43200.1000 | 43200.1000 |
| created: | 1461 | 1461 | 1461 | 1461 | 1461 |
| started: | 51790 | 67827 | 58011 | 63532 | 60312 |
| relayed: | 26212 | 39023 | 30856 | 36451 | 33768 |
| aborted: | 25577 | 28800 | 27154 | 27079 | 26541 |
| dropped: | 26262 | 38990 | 30821 | 37520 | 32479 |
| delivery_prob: | 0.2334 | 0.2981 | 0.1978 | 0.2822 | 0.2506 |
| delay_prob: | 0.2738 | 0.2370 | 0.2824 | 0.2480 | 0.2603 |
| hopcount_avg: | 3.8689 | 3.5145 | 4.3112 | 3.6429 | 3.7951 |
| buffertime_avg: | 1430.1069 | 1054.1910 | 1531.6268 | 1132.2647 | 1298.4176 |

opportunistic networks, particularly the message delay and delivery probability, when context information about user is available or not. The present findings clearly indicates that the context-based forwarding is a very interesting approach of communication in opportunistic networks, however, in comparison to flooding-based protocols it is not suitable. The present routing is able to give better result in presence as well as absence of context information, specifically in term of message delay and delivery probability.

Despite this, a number of directions exist in integrated routing which can be further investigated. For example we can improve performance of integrated routing in terms of message delay, message delivery, network congestion and resource consumption etc. Developing a network theory to model users' social relationships and exploit these models for designing routing protocols, this is a very interesting research direction in opportunistic network.

ACKNOWLEDGMENT

The authors are thankful to ABV-IIITM, Gwalior for the financial support, they have provided for carrying out present work. We are grateful to Prof. A. Trivedi, ABV-IIITM, Gwalior for his valuable interactions.

AUTHORS PROFILE

ANSHUL VERMA is a M.Tech (Digital Communication) student in the Computer Science and Engineering department at ABV-Indian Institute of Information Technology and Management, Gwalior, M.P., India. He has research interest in database management systems, Opportunistic Network, wireless communications and networking.

Dr. ANURAG SRIVASTAVA, working as an Associate Professor in Applied Science Departement at ABV-Indian Institute of Information Technology and Management, Gwalior, M.P., India. His Research activities are based on Model Calculation, Computational Physics/ Nano-Biophysics, Nano Electronics, IT Localization, E-Governance, IT Applications in different domains.